# Advection-Dominated Accretion Model of Sagittarius A* and Other Accreting Black Holes

Ramesh Narayan[1], Insu Yi[1,2], and Rohan Mahadevan[1]

[1] Harvard-Smithsonian Center for Astrophysics, 60 Garden St., Cambridge, MA 02138, USA
[2] Institute for Advanced Study, Olden Lane, Princeton, NJ 08540, USA

September 27, 1995

**Abstract.** Viscous rotating accretion flows around black holes become advection-dominated when the accretion rate $\dot M$ is sufficiently low. Most of the accretion energy in such flows is stored within the gas and advected radially inward. The temperature is therefore very high, and much of the radiation comes out in hard X-rays and $\gamma$-rays. We have constructed an advection-dominated accretion flow model for the Galactic Center source Sagittarius A*. The model consists of a $7 \times 10^5 M_\odot$ black hole accreting at $\dot M = 1.2 \times 10^{-5} \alpha \, M_\odot \, {\rm yr}^{-1}$, where $\alpha$ is the usual viscosity parameter. The model spectrum fits the observations from radio to $\gamma$-rays quite well and explains the unusually low luminosity of the source. Since the model explicitly makes use of a horizon at the inner edge to swallow the advected energy, the success of the model strongly suggests that the central object in Sgr A* is a black hole. We further show that, if $\alpha$ is not much smaller than unity, then advection-dominated models can be applied even to higher luminosity black holes. The existence of Low and High States in black hole X-ray binaries, and the abrupt transition between the two states, find a natural explanation. The models also explain the close similarity in the hard X-ray/$\gamma$-ray spectra of black hole X-ray binaries and active galactic nuclei.

## 1. Introduction

In a series of recent papers (Narayan & Yi 1994, 1995a,b, Abramowicz et al. 1995, Chen 1995, Chen et al. 1995), a class of accretion flow solutions has been discussed which appears to be relevant to accreting black holes and neutron stars. These solutions exist at relatively low mass accretion rates, $\dot m \lesssim 0.3\alpha^2$, where $\dot m$ is expressed in Eddington units (using a fiducial efficiency of 0.1) and $\alpha$ is the usual viscosity parameter.

The most important feature of the new flows is that the accreting gas is optically thin and has low radiative efficiency. Therefore, the bulk of the energy released through viscosity is advected radially inward with the gas as thermal energy and only a fraction of the energy is radiated. These *advection-dominated* solutions are very different from the usual cooling-dominated solutions considered in the past, where all the released energy is assumed to be radiated. The solutions are related to the "ion torus" model developed by Rees et al. (1982), but are distinct from another optically thick branch of advection-dominated solutions discovered by Abramowicz et al. (1988).

The new optically-thin advection-dominated branch of solutions has several very interesting properties which have been explored in the papers listed above. Two features of particular interest are the following:

(1) Because all the energy goes into internal energy, the gas becomes extremely hot. Indeed, Narayan & Yi (1995b) constructed two-temperature models (based on the ideas of Shapiro, Lightman & Eardley 1976) where the ions achieve nearly virial temperature, $T_i \sim 10^{12} {\rm K}/r$, and the electrons are cooler, $T_e \sim 10^9 - 10^{10}$ K for $r \lesssim 10^3$; here, $r$ is the radius expressed in Schwarzschild units. The nearly virial ion temperature causes the accreting gas to rotate with a sub-Keplerian angular velocity, and to take up a nearly spherical morphology rather than a disk-like or toroidal morphology (Narayan & Yi 1995a). The relativistic electron temperature makes it natural for the gas to radiate a large fraction of its luminosity in soft $\gamma$-rays.

(2) The flow is thermally stable (Abramowicz et al. 1995, Narayan & Yi 1995b) to local large wavelength perturbations. This is an important feature because prior to this the only hot accretion flow solution known was the thermally unstable solution discovered by Shapiro, Lightman & Eardley (1976). Kato, Abramowicz & Chen (1995) have recently shown that the new solutions may be weakly unstable to short wavelength thermal perturbations, but the instability does not appear to pose any danger to the overall viability of the flows. The solutions are also convectively unstable (Narayan & Yi 1994, 1995a, Igumenshchev, Abramowicz & Chen 1995), but again this instability does not threaten the survival of the flow. In fact, convection may help by enhancing turbulent viscosity, thereby increasing the value of $\alpha$.

The high temperature and thermal stability of the new solutions make them especially promising for explaining the hard X-ray and soft $\gamma$-ray spectra of accreting black hole systems. The advection-dominated nature of the flow suggests that the best applications are likely to be to low-luminosity systems. We describe one such application in §2, and discuss the extension of these models to higher luminosity systems in §3.

## 2. Application to Sgr A*

We have successfully applied the optically-thin advection-dominated accretion model to several low-luminosity systems: (i) the Galactic Center source Sagittarius A* (Sgr A*, Narayan, Yi & Mahadevan 1995), (ii) the nucleus of the LINER NGC 4258 (Lasota et al. 1995), and (iii) black hole soft X-ray transients in quiescence (Narayan, McClintock & Yi 1996, Yi et al. 1996). Here we discuss our model of Sgr A*.

The peculiar radio source Sgr A* at the Galactic Center has for many years been interpreted as a supermassive black hole ($M \gtrsim 10^6 M_\odot$) based on the dynamics of surrounding gas and stars (Genzel & Townes 1987, Genzel, Hollenbach & Townes 1994, Genzel 1996). The source has a luminosity of $\sim 10^{37}$ ergs s$^{-1}$, extending in a nearly flat spectrum (when represented as $\nu L_\nu$ vs $\nu$, see Fig. 1) from radio/mm through infrared up to hard X-rays. It is impossible to fit the unusual spectrum of the source with a standard thin accretion disk model (e.g. see the dashed line in Fig. 1). Furthermore, there is evidence for considerable gas flows in the vicinity of Sgr A* and it is estimated that the mass accretion rate on to the source, if it is a supermassive black hole, should be $\sim 10^{-4} M_\odot \text{yr}^{-1}$ (Melia 1992, Genzel et al. 1994). Such a large $\dot{M}$ would give an accretion luminosity of $\sim 10^{42}$ ergs s$^{-1}$, if the radiative efficiency is 10% (as expected with a thin disk model), which is clearly inconsistent with the observations. Various novel models have been proposed for Sgr A* (e.g. Falcke, Mannheim & Biermann 1993, Duschl & Lesch 1994, Melia 1994), but no model has yet been able to reconcile satisfactorily the requirements of low luminosity, flat spectrum and high mass accretion rate. We have constructed an advection-dominated model of Sgr A* in which a $7 \times 10^5 M_\odot$ black hole accretes at a rate of $1.2 \times 10^{-5} \alpha\, M_\odot \text{yr}^{-1}$ (Narayan, Mahadevan & Yi 1995, see Rees 1982 for a qualitative outline of the chief features of the model). We calculate the spectrum of the optically thin flow by including cyclo-synchrotron emission (Mahadevan, Narayan & Yi 1996), bremsstrahlung, and Comptonization. As Fig. 1 shows, the calculated spectrum gives quite a good fit to the observations. The independently fitted black hole mass falls close to the value suggested by the dynamical evidence (Genzel et al. 1994). More importantly, the mass accretion rate is quite high.

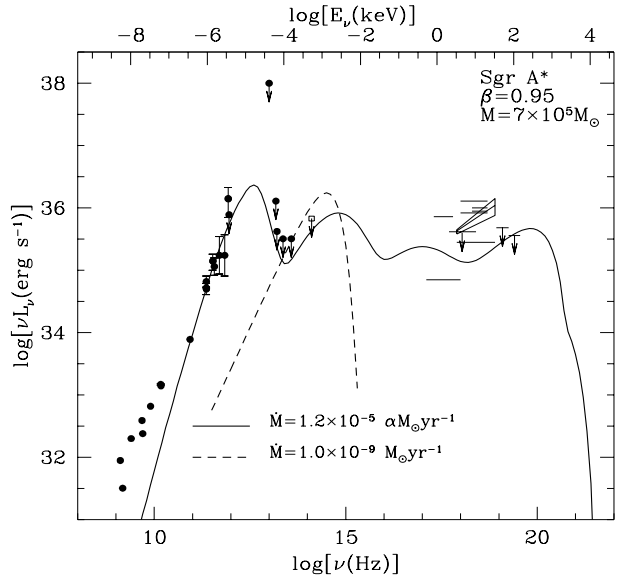

**Fig. 1.** The solid line shows the spectrum of Sgr A* corresponding to an advection-dominated accretion model with a black hole mass of $7 \times 10^5 M_\odot$ and a mass accretion rate of $\dot{M} = 1.2 \times 10^{-5} \alpha\, M_\odot \text{yr}^{-1}$. The model assumes that gas pressure accounts for 95% of the pressure ($\beta = 0.95$ in the notation of Narayan & Yi 1995b), with magnetic pressure accounting for the rest. The model gives a good fit to measurements over the entire spectral range from radio to $\gamma$-rays (see Narayan et al. 1995 for more details on the observational data). The dashed line is a standard thin disk model where the black hole mass has been kept the same and $\dot{M}$ has been reduced to $10^{-9} M_\odot \text{yr}^{-1}$ in order to satisfy the infrared upper limits. This model does not fit either the radio or the X-ray/$\gamma$-ray data. In addition, the $\dot{M}$ required in this model is far lower than the estimated $\dot{M}$ based on observations of gas in the vicinity of Sgr A*.

The last feature is a significant improvement over most previous models. Given the observed luminosity of the source, a standard cooling-dominated model would predict an accretion rate of only $\sim 10^{-9} M_\odot \text{yr}^{-1}$, which is in severe conflict with direct estimates of $\dot{M}$. The chief reason our model is able to employ a fairly substantial $\dot{M}$ and yet explain the very low luminosity of the source is that the model is advection-dominated (Rees 1982). In our model, more than 99.9% of the viscously dissipated energy is advected with the flow and disappears through the horizon of the black hole, and less than 0.1% of the energy is actually radiated.

This brings up the following important point. If the central star in Sgr A* were not a black hole but a normal object with a surface, say a $10^6 M_\odot$ "supermassive star," then our model would fail because the energy would ultimately be re-radiated from the stellar surface and would be seen somewhere in the spectrum. The success of our model depends critically on the assumption that the central object possesses a horizon. The model therefore provides strong confirmation that Sgr A* is indeed a black hole, as has long been suspected but never proved.

## 3. Application to High Luminosity Sources

The advection-dominated nature of the solutions makes them an obvious choice for modeling low-luminosity sources. However, the majority of black hole X-ray binaries (XRBs) and active galactic nuclei (AGNs) observed have luminosities $\gtrsim 0.01$ times the Eddington luminosity $L_{\rm Edd}$. Can the same models be applied also to such sources? We find that we can indeed use our solutions to model these sources provided we take $\alpha$ to be large.

Figure 2 shows a sequence of model spectra of a $10 M_\odot$ black hole. In these models, we assume that a standard thin disk extends from a large outer radius $r_{\rm out}$ down to a transition radius $r_{\rm tr}$, and that the flow then becomes hot and advection-dominated from $r_{\rm tr}$ down to $r_{\rm in} = 3$. Inside $r_{\rm in} = 3$, we assume that the radial velocity becomes supersonic and the emission is negligible. We have set $\alpha = 1$ in the hot inner flow. The different models shown in Fig. 2 correspond to different mass accretion rates $\dot m$ and transition radii $r_{\rm tr}$, as explained in the figure caption. We see several interesting features: 1. The solutions with $\dot m \sim 10^{-1.5} - 10^{-1}$ have very hard spectra with photon indices $\alpha_N \sim 1.6 - 2$ and with hard X-ray/soft $\gamma$-ray luminosities of $\sim 10^{36} - 10^{37.5}$ ergs s$^{-1}$. The spectra have thermal cutoffs at photon energies $E \sim$ few $\times 100$ keV. These features are very similar to the spectra of black hole XRBs in their so-called "Low State" (e.g. Grebenev et al. 1993, van der Klis 1994), and we propose that the Low State corresponds to such advection-dominated flows.

2. With decreasing $\dot m$, the luminosity $L$ decreases very rapidly, roughly as $L \propto \dot m^2$. This means that relatively small changes in $\dot m$ can lead to large variations in $L$, and may perhaps explain the large variability of the hard X-ray flux of black hole sources. Further we find that the electron temperature increases with decreasing $\dot m$, which is consistent with the indication in some X-ray transients such as GRO J0422+32 (J. Kurfess, private communication). Also, it is possible to achieve extremely low X-ray luminosities at the level of $\sim 10^{31} - 10^{34}$ ergs s$^{-1}$ with only moderate decrease in $\dot m$ down to $\sim 10^{-3}$. This provides a natural explanation for the "Quiescent/Off State" of the soft X-ray transients (Narayan, McClintock & Yi 1995, Yi et al. 1996). In our model, the Off State does not represent a decrease in $\dot m$ of several orders of magnitude, but rather a decrease of only one or at most two orders of magnitude relative to the Low State.

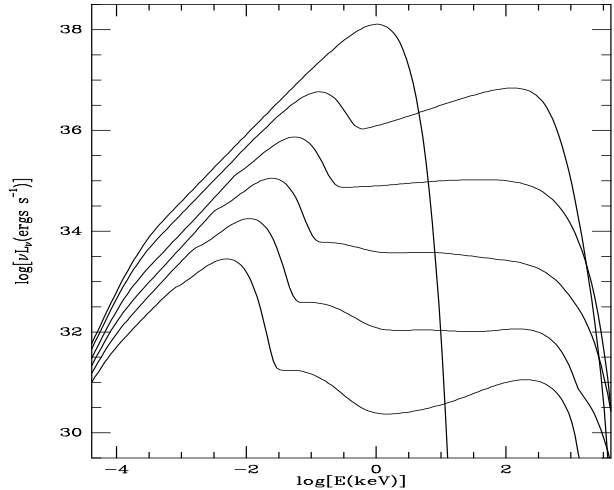

**Fig. 2.** Fig. 2. A sequence of model spectra for a $10 M_\odot$ black hole with $\alpha = 1$ and $\beta = 0.95$ (95% gas pressure). Starting from below, the models have $(\log \dot m, \log r_{\rm tr}) = (-3, 3.2), (-2.5, 2.9), (-2, 2.6), (-1.5, 2.3), (-1, 2),$ and $(-0.5, 0.5)$ respectively. Note that the luminosity increases very rapidly with increasing $\dot m$, and that for $\dot m \lesssim 0.1$ the spectra are quite hard. The bumps at the left end of the spectra are due to blackbody radiation from the cool outer disk. Above $\dot m \sim 0.1$, an optically-thin advection-dominated solution is no longer possible and so for $\log \dot m = -0.5$ we show the spectrum of a standard thin disk extending down to $r = 3$. This spectrum is very soft, cutting off above a few keV.

3. We find that the advection-dominated solution ceases to exist above a critical $\dot m \sim 0.1 - 0.2$ for $\alpha = 1$ (Narayan & Yi 1995b, Chen et al. 1995). Therefore, at high $\dot m$, the only solution we can construct is a thin disk with a soft spectrum (see the case corresponding to $\log \dot m = -0.5$ in Fig. 2). This naturally explains the sudden appearance in luminous black hole systems of the so-called "High State" with an ultra-soft spectrum. The transition between the Low State and the High State occurs at $\dot m \sim 0.3 \alpha^2$ in our models. Translated into luminosity, the maximum luminosity we can expect in the Low State is $\sim 0.05 \alpha^2 L_{\rm Edd}$. Since the observed maximum is about $(0.01 - 0.1) L_{\rm Edd}$ (Grebenev et al. 1993, van der Klis 1994), our models need a large value of $\alpha \sim 1$ to fit the observations. Even in the High State, black hole XRBs are known to have a low-luminosity hard tail in their spectra. This is probably due to a corona above the thin disk, which is not included in our models.

4. We have compared advection-dominated accretion flows around black holes and neutron stars. This is discussed in the accompanying paper by Yi et al. (1995).

5. We find that the shapes of advection-dominated model spectra are essentially independent of the mass of the black hole. Figure 3 shows a sequence of models ranging from $M = 10 M_\odot$, appropriate to XRBs, to $M = 10^9 M_\odot$, appropriate to the most powerful AGNs. In all cases, the spectra have $\alpha_N \lesssim 2$ with a cutoff at $\sim$ few $\times 100$ keV. This similarity of spectra has been noted in the observations, and is explained naturally by the models.

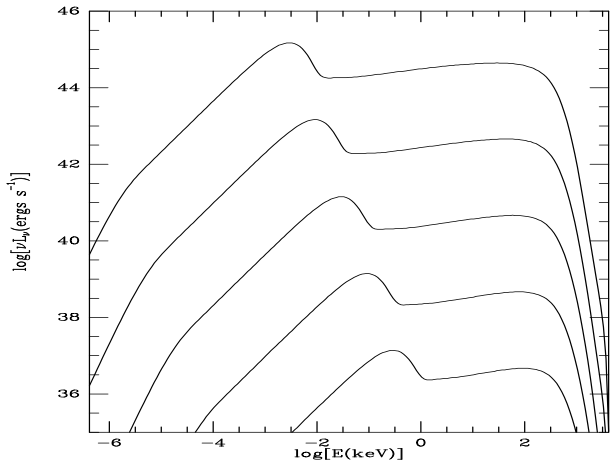

**Fig. 3.** A sequence of model spectra corresponding to increasing black hole mass upward. All the models have $\alpha = 0.1$, $\beta = 0.95$, $\dot{m} = 0.1$ and $\log r_{\rm tr} = 1.5$. Starting from below, the models correspond to black hole masses of $m = 10$, $10^3$, $10^5$, $10^7$, and $10^9$ respectively. Although the soft component in the spectrum (the bump) evolves from soft X-rays to UV as $m$ increases from 10 to $10^9$, the hard component is essentially independent of black hole mass. The models thus explain the similarity in hard X-ray/soft $\gamma$-ray spectra of black hole X-ray binaries and AGNs.

## 4. Summary

The optically-thin advection-dominated accretion solution is a *stable* and *hot* accretion flow model which naturally produces spectra with substantial luminosity in hard X-rays and soft $\gamma$-rays (of up to few $\times$ 100 keV). The model is therefore particularly well-suited for explaining the extensive observations of accreting black holes being carried out with the Compton Gamma-Ray Observatory.

The solutions work particularly well for low-luminosity sources like Sgr A$^*$ and quiescent soft X-ray transients. In addition, we find that the models work well even for luminous sources, provided we set $\alpha \sim 1$.

The model in its present form cannot explain hard power-law tails which extend to $E \gtrsim 1$ MeV. Such emission could arise from a non-thermal distribution of electrons which could in principle be included as an additional component in the model. The models developed so far, however, have considered only purely thermal electron distributions. Also, the variability properties of these models have not yet been considered, and it is not clear if observations like flicker and QPO can be explained naturally.

*Acknowledgement:* This work was supported in part by NSF grant AST 9423209. RN thanks the Inst. for Theor. Phys., Univ. of California, for hospitality.